\documentclass[epj]{svjour}

\usepackage{amsmath}
\usepackage{amssymb}
\usepackage{graphicx}

\newcommand{\bs}{\boldsymbol}

\newcommand{\mr}{\mathrm}

\usepackage{color}

\begin{document}

\title{Simulations of Sisyphus cooling including multiple excited states}

\author{F.\ Svensson \inst{1}, S.\ Jonsell\inst{2}, C.\ M.\ Dion\inst{1} }

\institute{\inst{1}Department of Physics, Ume{\aa} University, SE-901 87,
  Ume\aa, Sweden,\\ 
  \inst{2}Department of Physics, Swansea University, Swansea SA2 8PP, U.K., 
\email{b.s.jonsell@swansea.ac.uk}.}

   \abstract{ We extend the theory for laser cooling in a near-resonant
  optical lattice to include multiple excited hyperfine
  states. Simulations are performed treating the external degrees of
  freedom of the atom, i.e., position and momentum, classically, while
  the internal atomic states are treated quantum mechanically,
  allowing for arbitrary superpositions.  Whereas theoretical
  treatments including only a single excited hyperfine state predict
  that the temperature should be a function of lattice depth only,
  except close to resonance, experiments have shown that
  the minimum temperature achieved depends also on the detuning from
  resonance of the lattice light.  Our results resolve this
  discrepancy.}


\maketitle

\section{Introduction}

Laser cooling is a generic name for a number of techniques that use
laser light to cool atoms down to millikelvin or even microkelvin
temperatures \cite{gulaboken}. Today, laser cooling is used in
numerous applications, for instance as one of the steps used in the
process to create a Bose-Einstein
condensate \cite{pethick}, in atomic clocks \cite{wyn05}, and other
high-precision experiments using atoms.

The most commonly used technique, Doppler cooling, has as its lower
limit the Doppler temperature, which for most atoms is of the order
0.1 mK \cite{gulaboken}. However, lower temperatures can be reached in
near-resonant optical lattices, i.e., standing waves of laser light
with periodical, spatially alternating
polarisations~\cite{jes96,gry01}. In these systems temperatures
approaching the recoil limit of a few $\mu$K can be achieved
\cite{let88}. This result first came as a surprise, but was soon given
a theoretical model in form of the so-called Sisyphus mechanism
\cite{dal89,ung89}. In this model a combination of spatially dependent
optical pumping rates between the magnetic sublevels of the atom,
together with the periodic potentials of the optical lattice, give
rise to an effective friction which causes cooling of the atoms.

Whereas the Sisyphus model successfully describes many of the
qualitative features of laser cooling in optical lattices, it is not
sufficient for a quantitative analysis. For this purpose a number of
numerical techniques have been developed, e.g., semiclassical methods
based on Fokker-Planck-like equations \cite{pet99,jon06}, a
band-structure model \cite{cas91}, and quantum Monte Carlo simulations
\cite{dal92}. (For a review see reference~\cite{gry01}.) These theoretical
techniques have gone a long way in reproducing experimental
findings. However, unexplained features still remain. For instance,
theoretical simulations have consistently given kinetic temperatures
of the atoms which are independent of the laser detuning from the
atomic resonance, except very close to this
  resonance~\cite{san02}. Experimental results show that this is
largely true for large potential depths (large laser irradiances),
where the kinetic temperature is a linear function of the potential
depth.  However, as the laser irradiance is lowered a minimum
temperature is achieved, before the temperature starts to rapidly
increase for even lower irradiances. The point of this minimum is
often referred to 
as {\em d\'{e}crochage}. Experiments have shown that, in contrast to
theoretical predictions, the point of d\'{e}crochage does depend on
detuning \cite{jer00,ell01,car01}.

Hitherto all theoretical simulations have used a simplified level
structure of the atom. It has been assumed that the cooling process
only depends on optical pumping via a single excited state. However,
the excited state is really a manifold of several closely-lying
hyperfine states. In caesium the excited state used has total angular
momentum $F_{\mr e}=5$. However, separation between the $F_{\mr e}=5$
and $F_{\mr e}=4$ excited states is only $48.1\Gamma$ ($\Gamma$ being
the natural linewidth), which is comparable to typical detunings in
experiments. A recent experiment showed that even for detunings very
close to the $F_{\mr e}=4$ state, the proximity to this state did not
seem to have any effect on the temperature \cite{ell01}. This seems to
underpin the assumption that this state can be neglected in
simulations. Nevertheless, it is still possible that the $F_{\mr e}=4$
state is important for the cooling process close to d\'{e}crochage. In
this paper this possibility is investigated using semiclassical
simulations including both the $F_{\mr e}=5$ and $F_{\mr e}=4$ states
of the hyperfine manifold.

\section{Method}

In an earlier publication we developed a novel semiclassical method
for Sisyphus cooling \cite{jon06}, and showed that this method gives
excellent agreement with the fully quantum-mechanical method
\cite{dal92}. In the semiclassical method the external degrees of
freedom, i.e., position and momentum, are treated as simultaneously
well-defined classical variables. The internal degree of freedom,
i.e., the magnetic substate, is on the other hand treated fully
quantum mechanically, allowing for arbitrary superpositions. In this
way we are able to generalize the simplified $F_{\mr g}=1/2\rightarrow
F_{\mr e}=3/2$ model to realistic angular momenta, $F_{\mr
  g}=4\rightarrow F_{\mr e}=5$ for Cs, while retaining an excellent
agreement with fully quantum-mechanical simulations.  This is in
contrast to, e.g., the treatment in reference \cite{pet99} where the
internal states were projected onto an adiabatic basis. Our method
automatically includes all couplings between adiabatic (or diabatic)
states, which were neglected in reference \cite{pet99}. Inclusion of
these couplings has been showed to be crucial for good agreement with
fully quantum-mechanical simulations \cite{jon06}.

Having established the validity of our method, we now continue onto more
detailed investigations of the cooling process.  We first extend our
method to include two excited hyperfine states.  The generalized
optical Bloch equations for an atom of mass $m$ with a ground state
${\mr g}$ and 
two excited states ${\mr e}1$ and ${\mr e}2$ are
\begin{equation}
\label{eq:master}
  \mr{i}\hbar \frac{\partial \sigma}{\partial t}=\left[\frac{p^2}{2m}+{\mr
      H}_{\mr A}+V_{\mr{AL}}
    ,\sigma\right]+\left.\frac{\mr{d}\sigma}{\mr{d}t}\right
|_{\mr{sp}}.
\end{equation}
Here $\sigma$ is the density matrix of the atom, including the ground
state ${\mr g}$ and both excited states ${\mr e1}$ and ${\mr e2}$,
\begin{equation}
  \label{eq:sigma}
  \sigma=\left(
\begin{matrix}
\sigma_{\mr{gg}} & \sigma_{\mr{ge1}} & \sigma_{\mr{ge2}} \\
\sigma_{\mr{e1g}} & \sigma_{\mr{e1e1}} & \sigma_{\mr{e1e2}} \\
\sigma_{\mr{e2g}} & \sigma_{\mr{e2e1}} & \sigma_{\mr{e2e2}} 
\end{matrix} \right),
\end{equation}
where each $\sigma_{ij}$ is a submatrix with rows and columns
corresponding to the different magnetic sublevels. The Hamiltonian
part of the evolution is determined by the kinetic term, the atomic
internal Hamiltonian ${\mr H}_{\mr A}$ and the atom-laser interaction
$V_{\mr{AL}}$. Setting the zero of energy at the ground state, the
atomic Hamiltonian is just
\begin{equation}
  \label{eq:HA}
  {\mr H}_{\mr A}=\left(
    \begin{matrix}
      0 & 0 & 0\\
0 & \hbar \omega_{F_{\mr e1}} & 0 \\
0 & 0 & \hbar \omega_{F_{\mr e2}}
    \end{matrix}
\right),
\end{equation}
with $\hbar \omega_{F_{\mr e}}$ the energy of the excited
  states.  The atom-laser interaction takes the form
\begin{equation}
  \label{eq:VAL}
  V_{\mr{AL}} = -\hbar\left(
    \begin{matrix}
      0 & G_{\mr e1}^\dagger({\bf r}) & G_{\mr e2}^\dagger({\bf r})\\
      G_{\mr e1}({\bf r}) & 0 & 0 \\
      G_{\mr e2}({\bf r}) & 0 & 0
    \end{matrix}
\right),
\end{equation}
where $G_{\mr e}$ represents the simultaneous absorption of a photon
and excitation of the atom in the excited level $\mr e$, while $G_{\mr
  e}^\dagger$ represents the inverse emission process. The matrix
elements $G_{\mr e}$ are products of the appropriate transition
dipoles ${\bf d}$ and the positive frequency component ${\bf E}^+$ of
the laser field
\begin{align}
  \label{eq:G}
  \hbar G_{\mr e}({\bf r})& ={\bf d} \cdot {\bf E}^+({\bf r}),\\
{\bf E}^+({\bf r}) & = E_0 {\bs \xi}({\bf r}),
\end{align}
with ${\bs \xi}$ the (position-dependent) polarisation
  vector.  In the basis of the magnetic substates, $M_{\mr e}=-F_{\mr
  e},-F_{\mr e}+1,\dots,F_{\mr e}$ (${\mr e}= {\mr e1}$ or ${\mr
  e2}$), the transition dipole is the product of a reduced matrix
element and a Clebsch-Gordan coefficient,
\begin{align}
  \label{eq:dq}
{\bf d}_{\mr e} & = \langle F_{\mr e} || d || F_{\mr g}\rangle
\hat{\bf d}_{\mr e}, \\
\hat{\bf d}_{\mr e} & = \hat d^1_{\mr e}{\bs\epsilon}_{+1} +
\hat d^0_{\mr e}{\bs\epsilon}_0+\hat d^{-1}_{\mr e} {\bs\epsilon}_{-1}, \\
\hat d^q_{\mr e}& =   \langle F_{\mr g} 1
  M_{\mr g} q| F_{\mr e} M_{\mr e} \rangle,
\end{align}
and ${\bs\epsilon}_q$ are the usual spherical polarisation vectors.
The polarisation vector ${\bs \xi}$ is chosen as the
 one-dimensional lin$\perp$lin laser configuration~\cite{gry01},
\begin{equation}
  \label{eq:xi}
  {\bs \xi}(z)=\cos(kz){\bs \epsilon}_{-1}-i\sin(kz){\bs \epsilon}_{+1},
\end{equation}
with $k$ the wave vector of the laser.  The final term in
equation (\ref{eq:master}) describes the transfer of populations and
damping of coherences due to spontaneous emission
\begin{equation}
  \label{eq:sp}
  \left. \frac{\mr{d}\sigma}{\mr{d}t}\right|_{\mr{sp}}=\left(
    \begin{matrix}
    \gamma_{\mr{gg}} & -\sigma_{\mr{ge1}}\Gamma_{\mr e1}/2 &
    -\sigma_{\mr{ge2}}\Gamma_{\mr e2}/2 \\ 
    -\sigma_{\mr{e1g}}\Gamma_{\mr e1}/2 &
    -\sigma_{\mr{e1e1}}\Gamma_{\mr e1} & 0 \\ 
-\sigma_{\mr{e2g}}\Gamma_{\mr e2}/2 & 0 &  -\sigma_{\mr{e2e2}}\Gamma_{\mr e2}
    \end{matrix}
\right),
\end{equation}
where $\Gamma_{\mr e}$ is the partial
width of the excited state ${\mr e}$ for decay to the ground state
${\mr g}$. (Where decay to hyperfine states other than ${\mr g}$ are
possible, most experimental set-ups include a repumper laser which
brings the atom back to the excited state.)
Here $\gamma_{\mr{gg}}$ includes both the recoil and the probabilities
of populating different ground states after a cycle of optical
pumping through either of the two excited states,
\begin{multline}
  \label{eq:Ggg}
  \gamma_{\mr{gg}}=\frac{3\Gamma_{\mr e1}}{8\pi}\int {\mr d}\Omega_{\bs
    \kappa} \sum_{{\bs \epsilon} \perp {\bs \kappa}}
  [\hat{\bf  d}_{\mr e1} \cdot {\bs\epsilon}]^\dagger
e^{-i{\bs\kappa}{\bf r}}\sigma_{\mr{e1e1}}e^{i{\bs\kappa}{\bf r}}
\hat{\bf  d}_{\mr e1} \cdot {\bs\epsilon}\\
+\frac{3\Gamma_{\mr e2}}{8\pi}\int {\mr d}\Omega_{\bs
    \kappa} \sum_{{\bs \epsilon} \perp {\bs \kappa}}
  [\hat{\bf  d}_{\mr e2} \cdot {\bs\epsilon}]^\dagger
e^{-i{\bs\kappa}{\bf r}}\sigma_{\mr{e2e2}}e^{i{\bs\kappa}{\bf r}}
\hat{\bf  d}_{\mr e2} \cdot {\bs\epsilon}.
\end{multline}

In most cases of interest, the irradiance of the lasers is
sufficiently low that the population of the excited states is much
smaller than the ground-state population, i.e., the transitions are
far from saturation.  This condition can be expressed in terms of the
saturation parameters $s_{\mr e1}$ and $s_{\mr e2}$ as
\begin{equation}
  \label{eq:s}
  s_{\mr e}=\frac{\Omega_{\mr e}^2/2}{\Delta_{\mr e}^2+\Gamma^2_{\mr
      e}/4}\ll 1, 
\end{equation}
where $\Omega_{\mr e}=-\langle F_{\mr e} || d || F_{\mr g} \rangle
E_0/\hbar$ is the Rabi frequency of the transition and
$\Delta_{\mr e}$ the detuning from the excited state.  In the low
saturation limit the excited states will 
rapidly adjust to any change in the ground state density matrix. For
time scales relevant to the evolution of the ground state, the excited
state density matrices can be expressed as functions of
$\sigma_{\mr{gg}}$. Through the process of adiabatic elimination of
the excited states we then arrive at an effective equation for
$\sigma_{\mr{gg}}$ only,
\begin{multline}
  \label{eq:sgg}
 \frac{\mr d \sigma_{\mr{gg}}}{{\mr d} t}=\frac{1}{{\mr i}\hbar}\left[\hat{\mr
      H}_{\mr{eff}},\sigma_{\mr{gg}} \right]
-\sum_{\mr e}\frac{\Gamma'_{\mr e}}{2}\left\{A_{\mr e}(z),\sigma_{\mr{
    gg}} \right\} \\ +\sum_{\mr e}
\frac{3\Gamma'_{\mr e}}{8\pi}\int{\mr d}\Omega_{\bs\kappa}\sum_{{\bs
    \epsilon}\perp{\bs \kappa}}{B^{\mr
    e}_{\bs\epsilon}}^\dagger(z)e^{-{\mr
    i}{\bs\kappa}\cdot{\bf r}}\sigma_{\mr{gg}}e^{{\mr
    i}{\bs\kappa}\cdot{\bf r}}B^{\mr e}_{\bs\epsilon}(z),
\end{multline}
where
\begin{equation}
  \label{eq:B}
  B^{\mr e}_{\bs \epsilon}(z)=
\left[\hat{\bf d}_{\mr e}^\dagger \cdot {\bs  \xi}^*(z)\right]
 \,\left[\hat{\bf  d}_{\mr e} \cdot {\bs\epsilon} \right].
\end{equation}
Here the effective Hamiltonian describing the conservative part of the
evolution is given by
\begin{equation}
  \label{eq:H0}
  \hat{\mr H}_{\mr{eff}}=\frac{\hat{\mr p}^2}{2m}+\sum_{\mr e}\hbar\Delta'_{\mr
    e}A_{\mr e}(z).
\end{equation}
The matrix
\begin{equation}
  \label{eq:A}
  A_{\mr e}(z)=\left[\hat{\bf d}_{\mr e}^\dagger \cdot {\bs  \xi}^*(z)\right]
 \,\left[\hat{\bf  d}_{\mr e} \cdot {\bs\xi}(z) \right]
\end{equation}
describes the coherent couplings and light shifts of the magnetic
substates. We have also used the conventional notation
\begin{equation}
  \label{eq:DpGp}
  \Delta'_{\mr e}=\frac{\Delta s_{\mr e}}{2}, \qquad   \Gamma'_{\mr
    e}=\frac{\Gamma_{\mr e} s_{\mr e}}{2}. 
\end{equation}

In our simulations a semiclassical approximation to equation
(\ref{eq:sgg}) is used. This approximation is derived by first
rewriting equation (\ref{eq:sgg}) in terms of the Wigner distribution,
which is then Taylor-expanded to second order in $p$. For more details
see \cite{jon06}. This results in a set of coupled Fokker-Planck-like
equations for the populations of and coherences between the magnetic
substates. Hence, while position and momentum are treated as classical
variables, the internal states of the atom are treated fully quantum
mechanically.  Finally, the evolution equations are converted into a
Langevin form. That is, each atom is assigned a time-dependent
position $z(t)$ and momentum $p(t)$ (see, e.g., \cite{risken}), which
follow the classical equations
\begin{align}
  \dot{ z} & = \frac{p}{m},  \label{eq:xdot}\\
  \dot{ p} & = f(t)+\eta(t). \label{eq:pdot}
\end{align}
Here, $f(t)$ is a conservative force and $\eta(t)$ is a diffusive
force with the properties
\begin{equation}
  \label{eq:eta}
  \langle \eta(t)\rangle=0,\quad \langle \eta(t)\eta(t')\rangle=2 D(t) \delta(t-t'),
\end{equation}
where $\langle\cdot\rangle$ stands for a time average and $D$ is a
diffusion coefficient. The forces are calculated as a trace over the
magnetic sublevels, where the internal state of an atom is represented
by a density matrix $w(t)$.  The force is given by
\begin{align}
  \label{eq:force}
  f(t)=&
-\sum_{\mr e}\hbar\Delta_{\mr e}' \mr{Tr}\left\{{A_{\mr
    e}}^\prime(z)w(t)\right\} \nonumber  \\  \nonumber
& -{\mr i}\sum_{\mr e}\frac{\Gamma_{\mr  e}'}{2}
\sum_{q=0,\pm1}\text{Tr}\big\{[B^{\mr e}_q(z){B^{\mr
    e}_q}^{\dagger\prime}(z)  \\ & \qquad  - 
 B_q^{\mr e \prime}(z){B_q^{\mr e}}^\dagger(z)]w(t)\big\}
 .
\end{align}
The first term above is the force arising from the second-order
light-shift potential, while the second term is the radiation pressure.
The diffusion coefficient is given by
\begin{align} 
  D(t)
  =&\sum_{\mr e}\frac{\Gamma_{\mr
      e}'\hbar^2k_{\mr R}^2}{5}\sum_{q=0,\pm1}\frac{1}{1+\delta_{q0}} 
  \text{Tr}\left\{B^{\mr e}_q(z){B^{\mr e}_q}^\dagger(z)w(t)\right\}
  \nonumber \\
& + \sum_{\mr e}\frac{\Gamma_{\mr e}'\hbar^2}{2}\sum_{q=0,\pm1}
  \text{Tr}\big\{{B^{\mr e}_q}'(z) {B^{\mr
      e}_q}^{\dagger\prime}(z) w(t)\big\}, 
\end{align}
with $k_{\mr R}$ the wave vector of the emitted photon (we
  neglect the difference in energy of the photons emitted from the two
  excited states).
The first term arises from the recoil from photons spontaneously
emitted in random directions, while the second term is connected to
fluctuations in the radiation pressure. The evolution equation
for the internal-state density matrix is
\begin{align}
  \label{eq:w} \nonumber
  \dot{w}(t) = & \sum_{\mr e}\bigg\{{\mr i}\Delta_{\mr e}'[w(t),A_{\mr
    e}(z)]-\frac{\Gamma_{\mr e}'}{2}\{w(t),A_{\mr e}(z)\} \\ & 
+\Gamma_{\mr e}'\sum_{q=0,\pm1}{B_q^{\mr e}}^\dagger(z)w(t)B_q^{\mr
  e}(z)\bigg\}.
\end{align}

Finally, we consider the specific case of the $F_{\mr g}=4$ and
$F_{\mr e1}=4$, $F_{\mr e2}=5$ states of caesium. Using that $\langle4
|| d ||4 \rangle=\sqrt{7/12}\langle5 || d|| 4\rangle$ \cite{steck} we
find that $\Omega_4=(7/12)\, \Omega_5$. We also have $\Gamma_5=\Gamma$,
where $\Gamma$ is the natural linewidth, while for the the excited
state $F_{\mr e}=4$ the partial width for decay to the $F_g=4$ ground
state is $(7/12)\,\Gamma$. The energy separation between the excited
states gives $\Delta_4=48.1 \Gamma+\Delta_5$. From these relations the
respective saturation parameters (\ref{eq:s}) and $\Gamma_{\mr e}'$,
$\Delta_{\mr e}'$, equations (\ref{eq:DpGp}), can be derived.

\section{Results}

We have performed simulations using 5000 atoms for detunings
$\Delta_5=-10\Gamma$, $-20\Gamma$, $-30\Gamma$, $-40\Gamma$ and a
range of different potential depths (directly
  proportional to the parameter $\Delta_5'$). In all simulations the
initial temperature was $10\ \mu$K and the time step was ${\rm
  d}t=0.001/\Gamma'_5$.  
The simulations were iterated until the second moment of the momentum
distribution had stabilized. It should, however, be noted that since
simulations are necessarily performed using a finite number of atoms the
moments of the momentum distribution will still fluctuate over time
irrespectively of the number of iterations. 
The typical size of these fluctuations were $\Delta \langle p^2
\rangle \lesssim p_{\rm R}^2$ (where $p_{\rm R}$ is the recoil
momentum), but grow for potential depths below d\'{e}crochage. 
Depending on detuning and potential depth the
number of iterations required for convergence varied between 100000
and 600000.  The stability of the results were checked with larger
numbers of iterations and with shorter step sizes. It was found that
the step size $0.01/\Gamma'_5$, which was used in reference
\cite{jon06}, while working well at $\Delta_5=-10\Gamma$ was too
coarse for larger detunings.

\begin{figure}[htbp]
  \centering
  \includegraphics*[width=9cm]{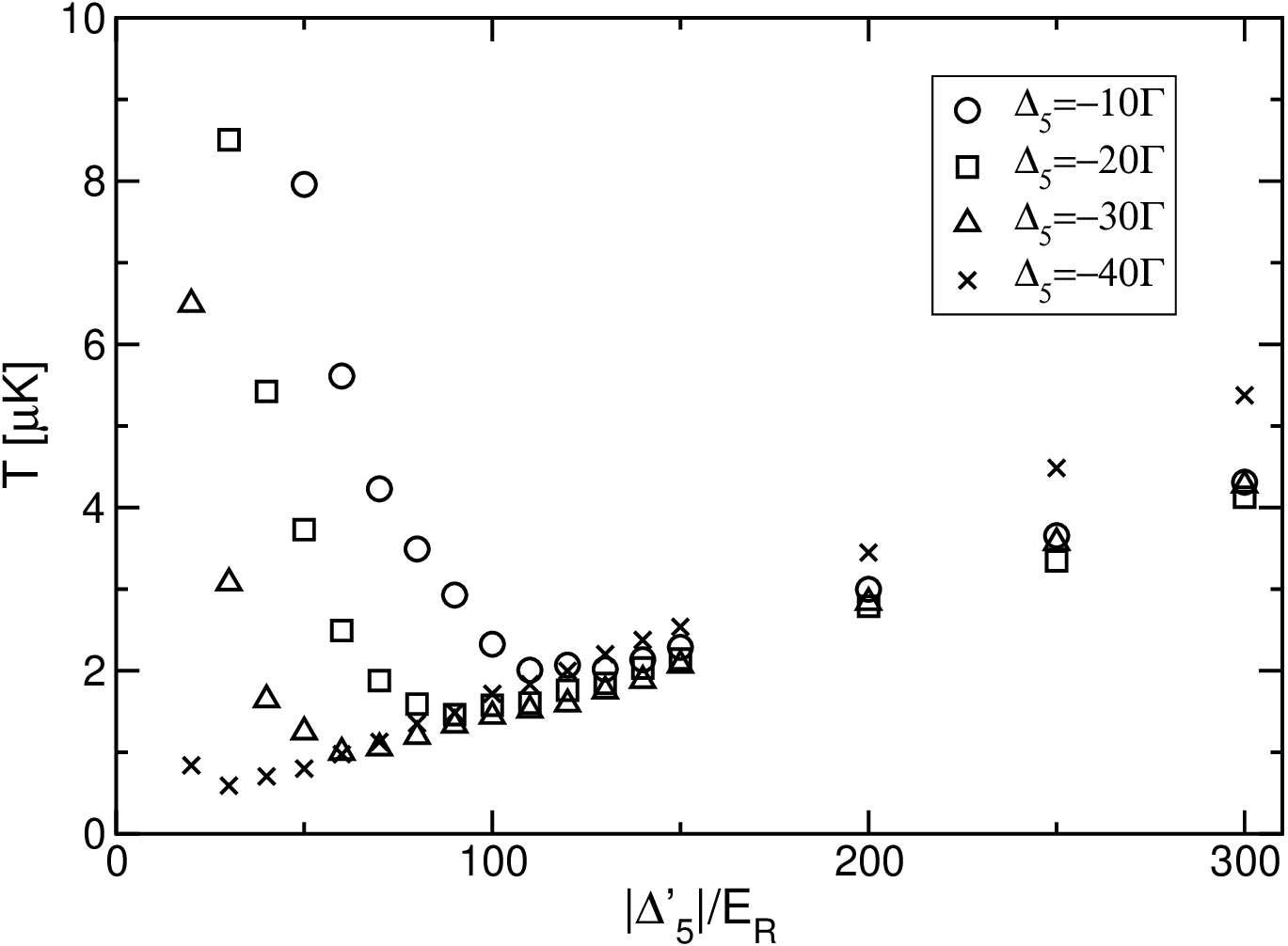}
  \caption{One-dimensional kinetic temperature as a function of
    potential depth for different detunings (see legend).  Simulations
    including both the $F_{\mr e}=4$ and $F_{\mr e}=5$ excited states. }
  \label{fig:graph1}
\end{figure}

Our main results are displayed in figure~\ref{fig:graph1}, where the
one-dimensional temperature $T$, defined as
$k_{\mr{B}}T/2=\langle p^2\rangle/2m$, where $k_{\mr{B}}$ is
  the Boltzmann constant and $m$ the atomic mass, is plotted against
$\Delta_5'$ for different detunings. 
The value of $\Delta_5'$ is expressed in terms of the recoil
energy gained by the atom after spontaneous emission, $E_{\mr R} =
\hbar^2 k_{\rm R}^2/2m$.
We note that for all detunings
except $\Delta_5=-40\Gamma$, the results for $|\Delta'_5| \gtrsim 125
E_{\mr R}$ fall on a single line. We thus confirm the experimental
finding in reference \cite{ell01} that the depth of the potential
generated by the $4\rightarrow5$ transition alone provides  the
appropriate scaling for large potential depths. The results for
$\Delta_5=-40\Gamma$ have a slightly different slope. Since for this
detuning the experimental data in reference \cite{ell01} only extends
up to $\Delta_5' \simeq 130E_{\mr R}$ it is not possible to say whether
this different slope is an experimental reality.

\begin{figure}[t]
  \centering
  \includegraphics*[width=9cm]{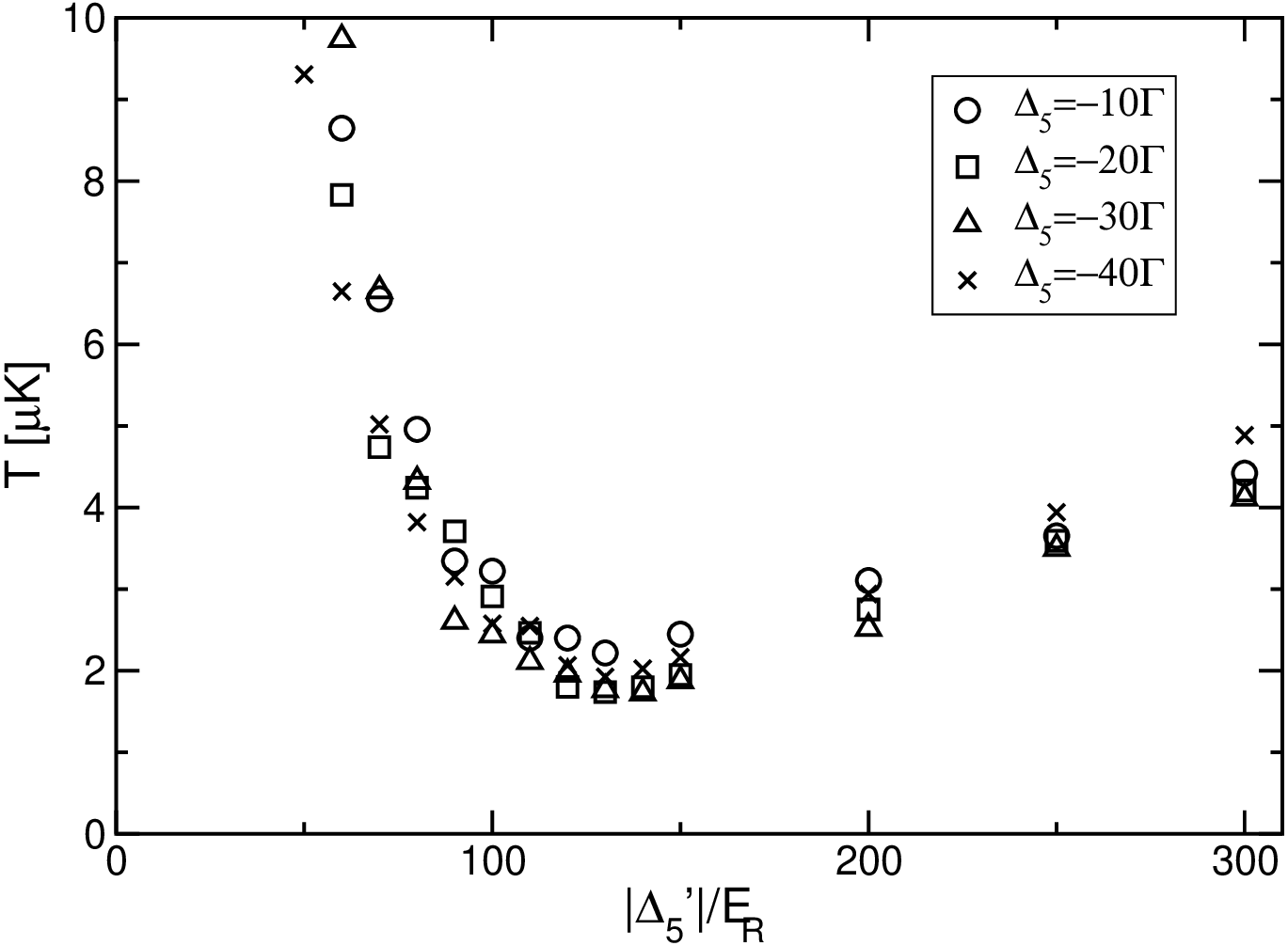}
  \caption{One-dimensional kinetic temperature as a function of
    potential depth for different detunings (see legend). Simulations
    including only the single $F_{\mr e}=5$ excited state.}
  \label{fig:graph2}
\end{figure}

In figure~\ref{fig:graph2} we compare to results of simulations
including only the single excited state $F_{\mr e}=5$. The most
striking difference is that in figure~\ref{fig:graph2} the results are
essentially independent of detuning over the whole range of potential
depths considered. When the additional excited state is included there
is a very clear dependence on detuning for shallow potentials, with
the linear dependence extending further for larger detunings, giving
rise to lower minimum temperatures.  For larger detunings the linear
behaviour is preserved to a lower potential depth, which makes it
possible to reach a lower temperature. This phenomenon has been
observed experimentally \cite{jer00,ell01,car01}, and has up to now
been at variance with all theoretical simulations, semiclassical or
fully quantum mechanical. Thus, we can conclude that it is the
additional excited state that causes this dependence of the point of
d\'{e}crochage on detuning.

\begin{figure}[htbp]
  \centering
    \includegraphics*[width=9cm]{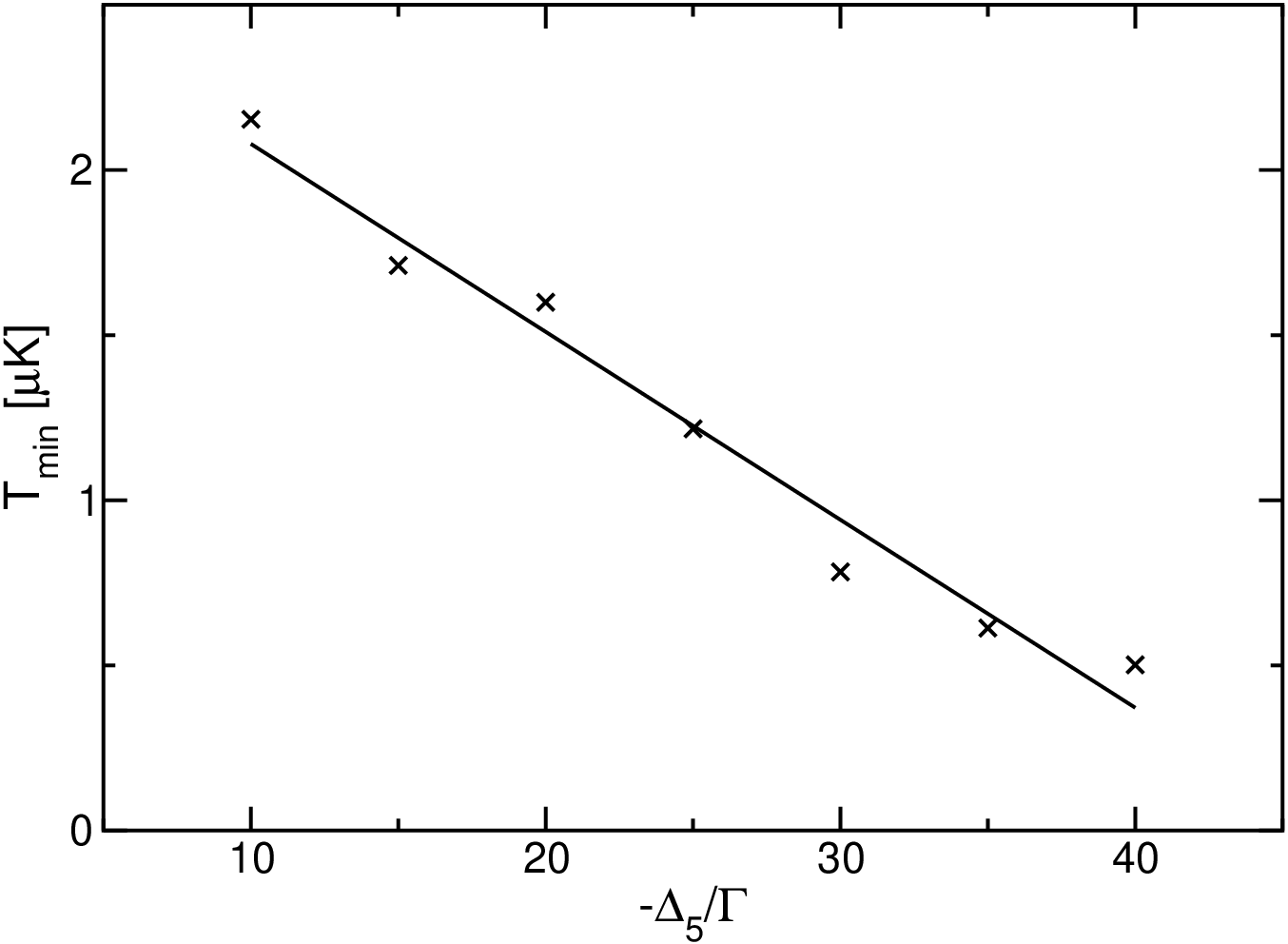}
  \caption{Minimum temperature $T_{\rm min}$ achieved for different detunings $\Delta_5$, together with a linear fit.}
  \label{fig:Tmin}
\end{figure}

\begin{figure}[htbp]
  \centering
    \includegraphics*[width=9cm]{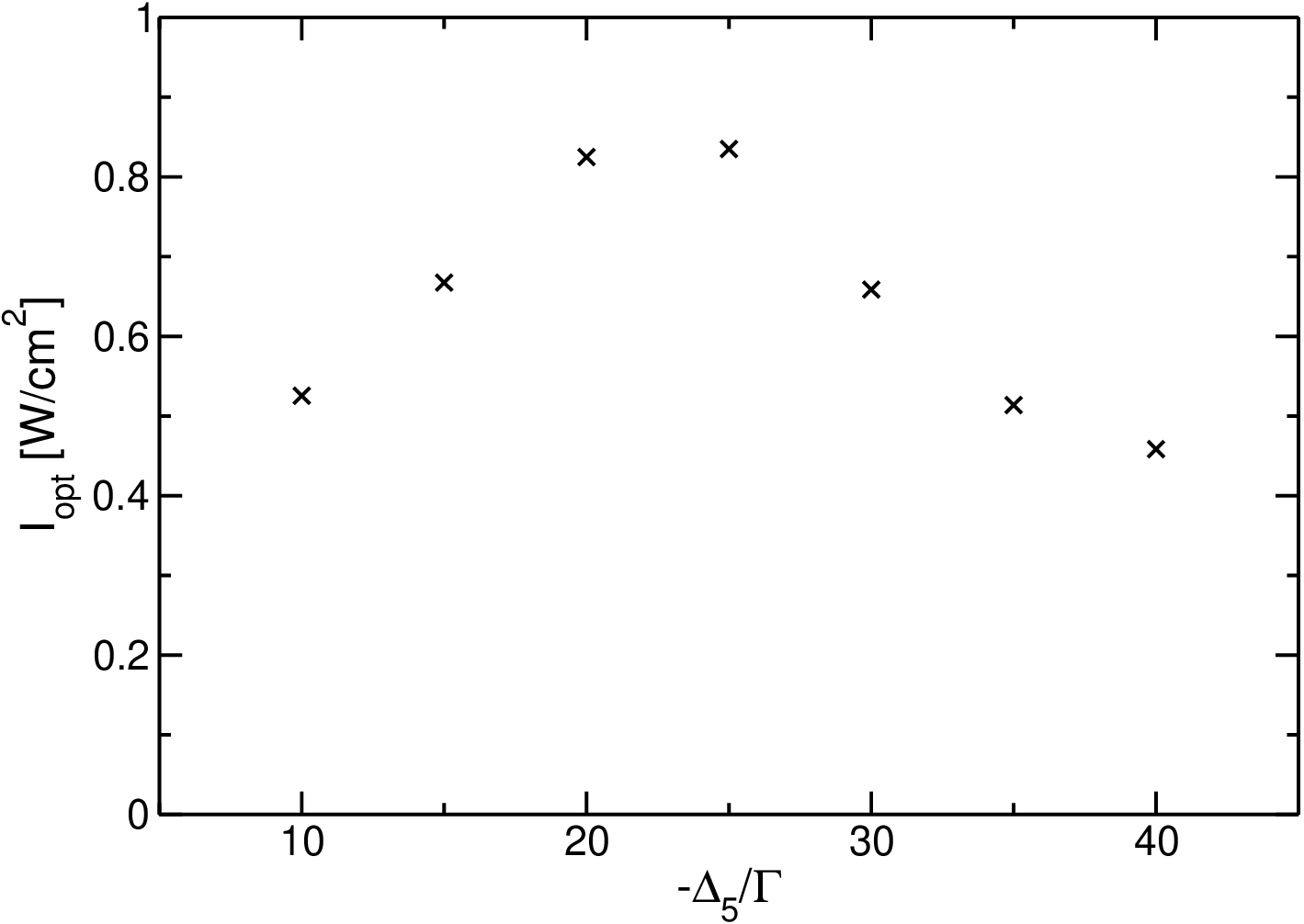}
  \caption{Laser irradiance $I_{\rm opt}$ for which the minimum temperature is
    achieved as a function of detuning $\Delta_5$.} 
  \label{fig:Iopt}
\end{figure}

In figure~\ref{fig:Tmin} we show the minimum temperature $T_{\rm min}$
achieved at different detunings, while figure~\ref{fig:Iopt} gives the
laser irradiance at which this minimum was obtained for the same
detunings. Around the minimum the simulations give a considerable
amount statistical noise. This is particularly true below
d\'{e}crochage, where a very small number of atoms with very 
high momenta have a significant impact on the value of $\langle p^2
\rangle$. In order to find the minimum we therefore made simulations
for $\Delta_5'$ in steps of $2E_{\rm R}$ around the minimum, and
fitted the results to the functional form 
$a |\Delta_5'| + b \exp(- c |\Delta_5'|)$,
with $a$, $b$ and $c$ fit parameters. It was found that this function
provides a good fit to simulated data to within the statistical
uncertainties. We find that both $T_{\rm min}$ and the optimal
potential depth follow a linear dependence on detuning. For $T_{\rm
  min}$ this is consistent with the results in \cite{jer00} over the
range of detunings considered. In \cite{jer00,car01} it was also
found that the minimum temperature is achieved for an optimal laser
irradiance ${\cal I}_{\rm opt}$  independent of detuning. 
Our results for ${\cal I}_{\rm opt}$ are displayed in figure
\ref{fig:Iopt}. The optimum irradiance varies almost a factor 2 over
the range of detunings investigated, and thus clearly deviates 
from the experimental result.

\section{Discussion}
\label{sec:disc}

At potential depths well above d\'{e}crochage, the $F_{\mr e}=4$ level has
very little influence. In reference \cite{ell01} it was reasoned that
for trapped atoms the dynamics is mainly determined by the lowest
adiabatic potential, as most atoms get optically pumped into
  the extreme $M_{\mr F}$ sublevels. This potential is not affected by the
$F_{\mr g}=4\rightarrow F_{\mr e}=4$ transition. If only the $F_{\mr
  g}=4\rightarrow F_{\mr e}=4$ transition is considered the lowest
adiabatic state has vanishing energy at all positions (for detunings
to the blue of the line), and is thus dark to the laser light. This
readily explains why, as long as the atoms are trapped in the lowest
adiabatic potential, their dynamics is determined only by the
parameters of the $F_{\mr g}=4\rightarrow F_{\mr e}=5$
transition. Even for large detunings the well known scaling of
temperature proportional to $I/\Delta_5 \propto \Delta_5'$ holds.

Recently the dynamics of laser cooling in
optical lattices was interpreted in terms of a bimodal momentum
distribution  \cite{san02,mar96,jer04,cla05}. The atoms are either in
an untrapped 
hot mode or in a 
cold mode where the atoms are trapped around a potential minimum. As
atoms are transferred from the hot to the cold mode the average kinetic
temperature decreases. When the system is in steady state the
interchange of atoms between the two modes is in balance. For large
potential depths essentially all atoms are trapped, giving rise to a
truncated Gaussian momentum profile. At lower potential
depths the hot mode can be observed even in steady state, giving rise
to a deviation from a Gaussian velocity profile in the wings of the
distribution.

\begin{figure}[htbp]
  \centering
  \includegraphics*[width=9cm]{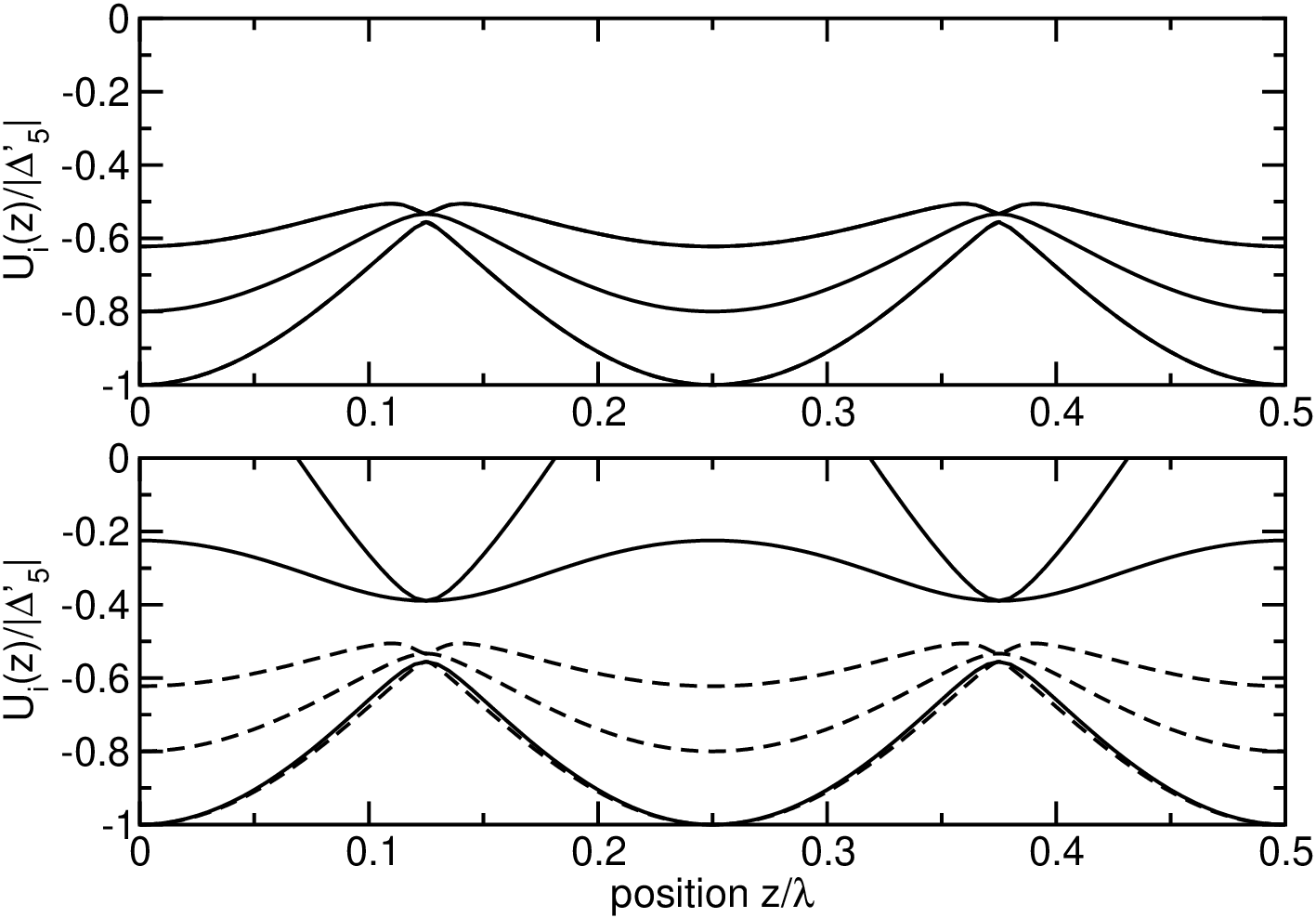}
  \caption{The three lowest adiabatic potentials $U_i(z)$ scaled by
    $|\Delta_5'|$ for
    $\Delta_5=-10\Gamma$ (upper panel) and $\Delta_5=-40\Gamma$ (lower
    panel). The solid line shows potentials calculated including both
    the $F_{\mr e}=4$ and $F_{\mr e}=5$ excited states, while the
    dashed line shows potentials calculated including the $F_{\mr
      e}=5$ excited state only. In the upper panel the solid and
    dashed lines are almost identical. The potentials including  only
    a single excited state scale with $\Delta_5'$ only, and are hence
    identical in the upper and lower panel.  }
  \label{fig:pot}
\end{figure}

The dependence of the point of d\'{e}crochage on detuning found in
this paper and in reference\ \cite{ell01} can also be understood from the
bimodal picture of Sisyphus cooling. 
The cold mode is, 
as explained above, completely determined by the $F_{\mr
  g}=4\rightarrow F_{\mr e}=5$ transition, and hence its temperature
scales proportionally to $\Delta_5'$ only. According to the bimodal model
 the hot mode starts to get populated around d\'{e}crochage,
 thus driving up the value of $\langle p^2 \rangle$,
even though most atoms are still trapped in the cold mode
\cite{jer04,cla05}. Even a 
relatively small population of the hot mode will dominate the value of
$\langle p^2 \rangle$ since the atoms in this mode have no upper limit
for their momenta, and for shallow potentials $\langle p^2 \rangle$
may even diverge \cite{lutz04}. 
The increased population of the hot mode is associated with atoms
leaving the lowest adiabatic state.
Our interpretation of the results in figure \ref{fig:graph1} is that
while the form of the 
cold mode is unaffected  when the $F_{\mr g}=4\rightarrow F_{\mr e}=4$
is taken into account, the rate of {\em transfer} of atoms from the cold to
the hot mode, i.e.\ away from the lowest adiabatic state, is reduced.  
As the magnitude of this effect depends on the 
laser detuning from  the $F_{\mr g}=4\rightarrow F_{\mr  e}=4$ transition
 a dependence of the point
of d\'{e}crochage on detuning is introduced in this way. 

 In figure \ref{fig:pot}  we show
the three lowest adiabatic potentials for detunings
$\Delta_5=-10\Gamma$ and $\Delta_5=-40\Gamma$.  As noted above the
lowest potential is identical for both detunings, giving the same
dynamics for both cases.  However, as the potential depth is reduced
the excited states in figure~\ref{fig:pot} gain significant
populations.  Since the potentials of these excited states are very
different at different detunings the universal temperature dependence
is violated.  The universal dependence persists to lower potential
depths at large detunings. We therefore conclude that the transfer of atoms
to adiabatic states with higher energies is more likely at small
detunings, while for large detunings the potential depth has to be
lowered even further before this transfer becomes important.  As shown
in figure~\ref{fig:pot} at $\Delta_5=-10\Gamma$ 
(and for potentials where the $F_{\rm g}=4\rightarrow F_{\rm e}=4$
transition has been excluded) 
there are avoided
crossings involving the lowest adiabatic potential at $z=\lambda/8$
and $z=3\lambda/8$ ($\lambda$ is the laser wavelength),
while for $\Delta_5=-40\Gamma$ there are distinct gaps. A tentative
conclusion is therefore that the formation of this gap inhibits
transfer from the cold to the hot mode, although the details of this
effect remains to be worked out.


\begin{figure}[htbp]
  \centering
  \includegraphics*[width=9cm]{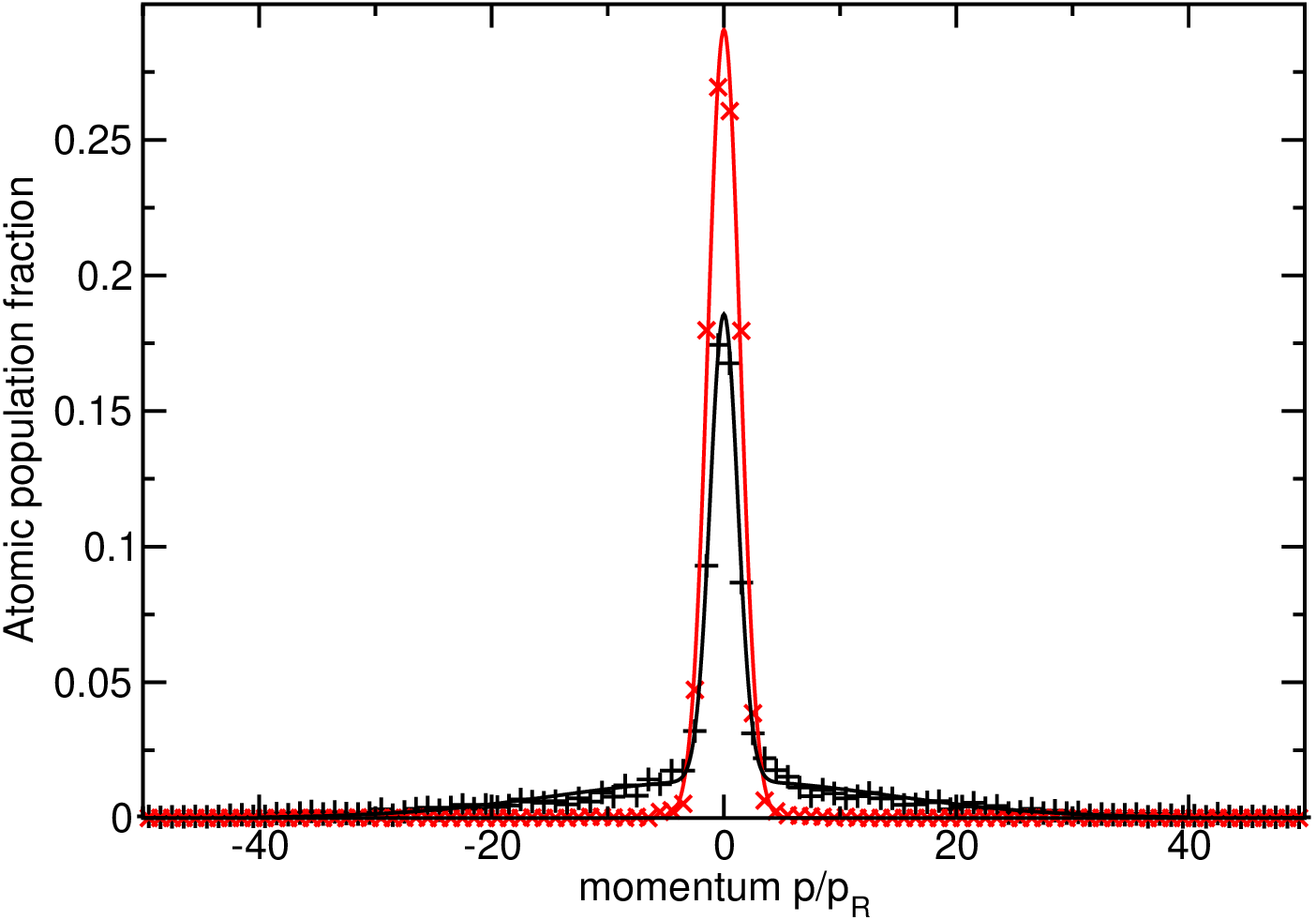}
  \caption{Momentum distributions for $\Delta_5=-10\Gamma$ (plus
    signs, black online) together with a fit to a double Gaussian, and
    for $\Delta_5=-40\Gamma$ (crosses, red online) together with a fit
    to a single Gaussian (double Gaussian would look the same). The
    momentum distribution has been binned into $1 p_{\mr R}$-wide
    bins, where $p_{\mr R} = \hbar k_{\mr R}$. The
  potential depth was $|\Delta_5'| = 20E_{\mr R}$, i.e., well below
  d\'{e}crochage for $\Delta_5=-10\Gamma$.}
  \label{fig:mom}
\end{figure}

This interpretation is also supported by the simulated momentum
profiles.  As an example momentum profiles at $|\Delta_5'|=20 E_{\mr
  R}$ at detunings $\Delta_5=-10\Gamma$ and $\Delta_5=-40\Gamma$ are
displayed in figure~\ref{fig:mom}, together with fits to double and
single Gaussians respectively. (Fitting also the $\Delta_5=-40\Gamma$
profile to a double Gaussian gives no improvement as the widths of the
two Gaussians in this case adjust to the same value, indicating that
the distribution really is well described by a single Gaussian.) For
the smaller detuning the wider hot mode is clearly visible. The fit
gives for the cold mode (i.e., the central peak) widths corresponding
to $\langle p^2 \rangle = 1.9p_{\mr R}^2$ for $\Delta_5=-40\Gamma$ and
$\langle p^2 \rangle = 1.4p_{\mr R}^2$ for
$\Delta_5=-10\Gamma$. Adding the hot mode gives a total $\langle p^2
\rangle = 142p_{\mr R}^2$ for $\Delta_5=-10\Gamma$, even though the
integral of the Gaussian reveal that both modes contain roughly the
same number of atoms (49\% hot, 51\% cold). Even at the larger
detuning a very small number of hot atoms increases $\langle p^2
\rangle$ to $4.1\,p_{\mr R}^2$. Considering the more than one order of
magnitude difference in the overall $\langle p^2 \rangle$ between the
two detunings, we find that the width of the cold mode is remarkably
similar, showing that indeed even well below d\'{e}crochage there is a
significant population of the cold mode, with characteristics largely
independent of the detuning.

\section{Conclusions}

In summary, we showed that at low potential depths, around the
so-called point of d\'{e}crochage, the temperature achieved by Sisyphus
cooling does depend on {\em both} the potential depth and the detuning
from resonance. For these potential depths it is necessary to include
several excited hyperfine state in the theoretical description, in
order to get accurate results. Simulations including only a single
excited hyperfine state show no dependence of the
temperature on detuning.  This finding agrees very well with the
experimental results of references
\cite{jer00,ell01,car01}, previously unreproduced by simulations. 
At larger potential depths, where the
temperature depends linearly on potential depth, we find that the
additional excited hyperfine state has no effect. This is also in
agreement with the experimental results in reference \cite{ell01} that
the temperature scales with the potential determined including {\em
  only} the $F_{\rm g}=4\rightarrow F_{\rm e}=5$ transition.

\begin{acknowledgement}

We thank Anders Kastberg, Stefan Petra, and Peder Sj\"{o}lund for many
valuable discussions.
This work was supported by the Swedish Research Council (VR) and by
the EPSRC through grant number  EP/D069785/1.

\end{acknowledgement}


\begin{thebibliography}{99}
\bibitem{gulaboken} H.~J.\ Metcalf, P.\ van der Straten, {\em
    Laser cooling and trapping} (Springer, New York, 1999)
\bibitem{pethick} C.~J.\ Pethick, H.\ Smith, {\em Bose-Einstein
    condensation in dilute gases} (Cambridge University Press,
  Cambridge, 2002)
\bibitem{wyn05} R.\ Wynards, W.\ Weyers, Metrologica {\bf 42}, S64
  (2005)
\bibitem{jes96} P.~S.\ Jessen, I.~H.\ Deutsch, Adv.\ At.\ Mol.\
  Opt.\ Phys.\ {\bf 37}, 95 (1996)
\bibitem{gry01} G.\ Grynberg, C.\ Robillard, Phys.\ Rep.\ {\bf
    355}, 335 (2001)
\bibitem{let88} P.\ Lett, R.\ Watts, C.\ Westbrook, W.D.\ Phillips,
  P.\ Gould, H.\ Metcalf, Phys.\ Rev.\ Lett.\ {\bf 61}, 169 (1988)
\bibitem{dal89} J.\ Dalibard, C.\ Cohen-Tannoudji, J.\ Opt.\ Soc.\
  Am.\ B {\bf 6}, 2023 (1989)
\bibitem{ung89} P.~J.\ Ungar, D.~S.\ Weiss, E.\ Riis, S.\ Chu,
  J.\ Opt.\ Soc.\ Am.\ B {\bf 6}, 2058 (1989)
\bibitem{pet99} K.~I.\ Petsas, G.\ Grynberg, J.-Y.\ Courtois,
  Eur.\ Phys.\ J.\ D {\bf 6}, 29 (1999)
\bibitem{jon06} S.\ Jonsell, C.~M.\ Dion, M.\ Nyl\'{e}n, S.~J.~H.\ Petra, P.\
  Sj\"{o}lund, A.\ Kastberg, Eur. Phys. J. D {\bf 39}, 3889 (2006)
\bibitem{cas91} Y.\ Castin, J.\ Dalibard, Europhys.\ Lett.\ {\bf
    14}, 761 (1991)
\bibitem{dal92} J. Dalibard, Y.\ Castin, K.\ M\o lmer, Phys.\ Rev.\
  Lett.\ {\bf 68}, 580 (1992)
\bibitem{san02} L.\ Sanchez-Palencia, P.\ Horak, G.\ Grynberg,
  Eur.\ Phys.\ J.\ D {\bf 18}, 353 (2002)
\bibitem{jer00} J.\ Jersblad, H.\ Ellmann, A.\ Kastberg, Phys.\
  Rev.\ A {\bf 62}, 051401 (R) (2000)
\bibitem{ell01} H.\ Ellmann, J.\ Jersblad, A.\ Kastberg, Eur.\
  Phys.\ J.\ D {\bf 13}, 379 (2001)
\bibitem{car01} F.-R.\ Carminati, M.\ Schiavoni, L.\ Sanchez-Palencia,
  F.\ Renzoni, G.\ Grynberg, Eur.\ Phys.\ J.\ D {\bf 17}, 249 (2001)
\bibitem{coh90} C.\ Cohen-Tannoudji, in 
{\it Fundamental systems in Quantum Optics},
Les Houches summer school of
  theoretical physics 1990, session {\bf LIII}, edited by J.\
  Dalibard, J.-M.\ Raimond, J.\ Zinn-Justin
(Elsevier Science Publishers, Amsterdam, 1992), p.1
\bibitem{risken} H.\ Risken, {\it The Fokker-Planck Equation}, 2nd
  edn. (Springer, Berlin, 1996) 
\bibitem{steck} D.~A.\ Steck, {\it Cesium D Line Data}, {\tt
    http://steck.us/alkalidata}
\bibitem{mar96} S.~Marksteiner, K.~Ellinger and P.~Zoller,
  Phys.~Rev.~A {\bf 53}, 3409 (1996)
\bibitem{jer04} J.~Jersblad, H.~Ellmann, K. St\o chkel, A.~Kastberg, L.~Sanchez-Palencia
and R.~Kaiser, Phys.~Rev.~A {\bf 69}, 013410 (2004);
\bibitem{cla05} C.~M.\ Dion, P.\ Sj\"{o}lund, S.~J.~H.~Petra, S.~Jonsell,
  A.\ Kastberg, Europhys. Lett. {\bf 72}, 369 (2005)
\bibitem{lutz04} E.\ Lutz, Phys.\ Rev.\ Lett.\ {\bf 93}, 190602 (2004) 
\end{thebibliography}
\end{document}